\documentclass[conference]{IEEEtran}
\IEEEoverridecommandlockouts

\usepackage{cite}
\usepackage{amsmath,amssymb,amsfonts}
\usepackage{algorithmic}
\usepackage{graphicx}
\usepackage{textcomp}
\usepackage{xcolor}
\usepackage{url}
\usepackage{hyperref}
\hypersetup{
    colorlinks=true,
    linkcolor=blue,
    citecolor=blue,
    urlcolor=blue,
    filecolor=blue
}
\usepackage{multirow}
\usepackage{stfloats}

\def\BibTeX{{\rm B\kern-.05em{\sc i\kern-.025em b}\kern-.08em
    T\kern-.1667em\lower.7ex\hbox{E}\kern-.125emX}}
\begin{document}

\title{Exposing and Mitigating Neural Codec Vulnerabilities in Audio Deepfake Detection
\\

\thanks{This work was supported by the National Science Foundation (NSF) under Grant No. 2409577.}
}

\author{\IEEEauthorblockN{Abdullah}
\IEEEauthorblockA{\textit{College of Innovation and Technology} \\
\textit{University of Michigan-Flint}\\
Michigan, USA \\
abdulahh@umich.edu}
\and
\IEEEauthorblockN{Awais Khan}
\IEEEauthorblockA{\textit{College of Innovation and Technology} \\
\textit{University of Michigan-Flint}\\
Michigan, USA \\
mawais@umich.edu}
\and
\IEEEauthorblockN{Khalid Mahmood Malik}
\IEEEauthorblockA{\textit{College of Innovation and Technology} \\
\textit{University of Michigan-Flint}\\
Michigan, USA \\
drmalik@umich.edu}

}

\maketitle

\begin{abstract}
Existing audio deepfake detection (ADD) datasets and detectors are primarily built for vocoder-based synthesis, evaluated against traditional post-hoc perturbations such as MP3/AAC compression or additive noise, applied independently of generation. However, recent speech synthesizers, particularly audio language model (ALM)-based systems, use neural audio codecs both for compression and as the resynthesis module reconstructing waveforms from generated tokens, producing artifacts distinct from post-hoc compression. Neural codecs thus play a dual role: some are designed for pure compression under low-bandwidth communication, while others serve as resynthesis components within ALM-based synthesizers. Despite this dual role, robustness to codec-based compression, unlike post-hoc compression, remains largely unexplored. We expose this gap, showing that state-of-the-art (SOTA) ADD models degrade drastically on codec-compressed speech; in particular, systems trained on Codec Resynthesized (CoRS) data as a proxy for codec-based generation prove most vulnerable, with legitimately compressed bona fide speech often misclassified as fake. To investigate this, we construct the Audio Neural Codec-Spoof (ANC-Spoof) dataset by applying seven neural codec algorithms to existing ADD benchmarks, isolating codec-induced resynthesis artifacts as a controlled proxy for codec-based generation. As baseline mitigation, we propose PCL-NET (Pairwise Consistency Learned Network), fine-tuning a pretrained XLS-R (300M) encoder with a pairwise consistency objective that minimizes the representation distance between an utterance's uncompressed and codec-compressed versions, disentangling codec artifacts from the real-versus-fake decision. As a result, PCL-NET reduces average EER under neural codec compression from 28.67\% (SOTA average) to 12.77\%, while preserving competitive CoSG-based deepfake detection performance. We will also make the dataset publicly available on Hugging Face upon acceptance.

\end{abstract}


\begin{figure}[htbp]
\centerline{\includegraphics[width=\columnwidth]{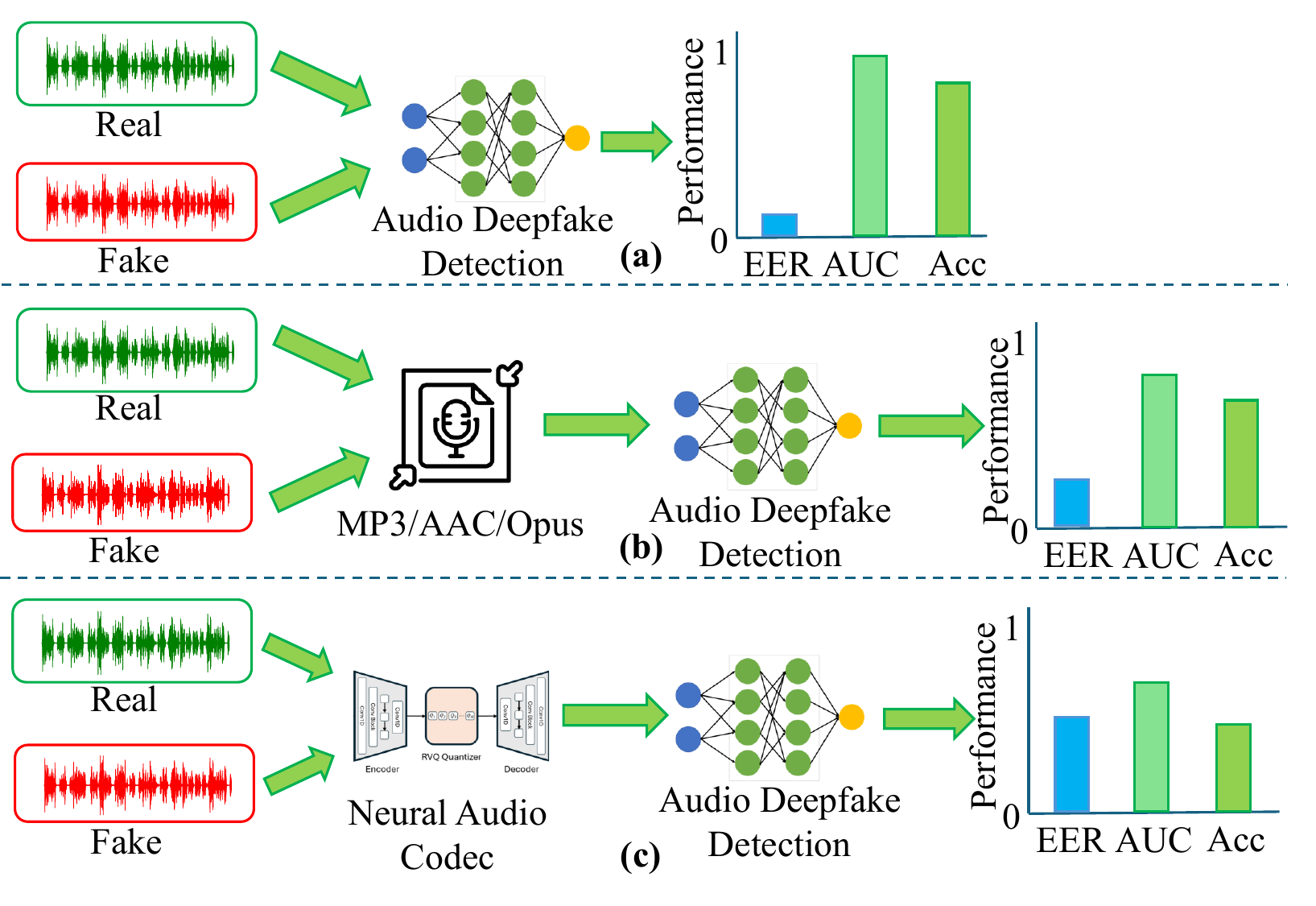}}
\vspace{-0.3cm}
\caption{Conceptual illustration of ADD performance under uncompressed, conventional, and neural codec compression. (a) Reliable detection of uncompressed fake audio (low EER, high AUC/accuracy). (b) Conventional compression (e.g., MP3/AAC/Opus) moderately degrades performance. (c) Neural audio codec compression causes more severe degradation.}
\vspace{-0.7cm}
\label{fig}
\end{figure}

\section{Introduction}
The growing sophistication of synthetic speech generation has made audio deepfakes a pressing concern across voice-based authentication, telecommunication security, and misinformation. To tackle this, audio deepfake detection (ADD) has matured into an active area of research~\cite{khan2023battling}, with substantial progress dedicated to improving generalization beyond the specific conditions observed during training. However, a large portion of this effort has centered on robustness to signal-level perturbations, such as additive noise, partial manipulations, and conventional, post-hoc compression codecs like MP3 or AAC \cite{wu2025clad, li2026measuring, 10447500, 10374968, khan2023spotnet}. These efforts have meaningfully improved ADD reliability under everyday transmission conditions as shown in Figure~\ref{fig} (a,b). Yet, they share a common assumption that has gone largely unexamined: that compression is a post-perturbation applied after speech has already generated, rather than a process that can be entangled with generation itself~\cite{uddin2025adversarialattacksaudiodeepfake,11375806}. \\
This assumption no longer holds for a growing class of speech synthesizers. Neural audio codecs, built on learned encoder-decoder architectures rather than traditional waveform or parametric schemes, have become foundational to modern speech technology. In one role, they compress efficiently, as in Google's Lyra codec, which enables intelligible speech over highly bandwidth-constrained networks and consumer devices \cite{kleijn2021generative,khan2026psa,10972633}. In a second, the neural codecs resynthesize speech within audio language model (ALM)-based text-to-speech and voice conversion systems: an autoregressive model predicts discrete tokens, and the codec decoder reconstructs the waveform from them \cite{borsos2023audiolm}. The codec decoder is therefore no longer a post-hoc stage, but part of generation itself, and its artifacts differ accordingly from those studied in prior robustness work.\\
This dual role introduces a subtle but consequential problem for ADD systems. Since most Codec Speech Generation (CoSG) systems remain proprietary, the community has adopted Codec Resynthesized (CoRS) speech, real audio passed through a codec's encoder-decoder without any generative modeling, as a training proxy, with reasoning that CoRS and CoSG share the same final decoding stage \cite{xie2025codecfake, chen2026codecfakeplus}. This assumption carries an unexamined risk: a detector that learns to associate codec distortion with the fake class may misflag legitimately compressed speech, such as audio from a voice-messaging app, simply because it shares the same decoding artifacts. Despite the ubiquity of neural codecs in everyday communication, systematic evaluation of ADD robustness under legitimate codec compression, and whether CoRS-based training worsens this vulnerability, remains largely absent (Figure~\ref{fig}c).
To address this gap, we first evaluate state-of-the-art (SOTA) ADD systems, standard, perturbation-robust, and CoRS-trained, under compression from seven diverse neural codecs. Our detailed analysis reveals that detection performance degrades sharply under codec compression even for otherwise-robust systems, with CoRS-trained systems proving most fragile due to their implicit reliance on codec artifacts as a discriminative cue. Motivated by this, we design PCL-NET (Pairwise Consistency Learned Network), a model with a novel training strategy that separates codec-induced distortion from the notion of fakeness. We apply neural codec compression to both bona fide and spoofed utterances during training and introduce a pairwise consistency objective that keeps each utterance's self-supervised representation stable across its original and compressed forms, encouraging codec-invariant rather than artifact-driven features.
The overall contributions of this work are as follows:
\begin{itemize}
\item We present the first systematic study exposing that existing ADD systems, including CoRS-trained variants, suffer substantially under neural codec compression, a distortion fundamentally different from the conventional perturbations; we further show that CoRS-based systems exacerbate this vulnerability and increase the misclassification of legitimately compressed real speech as fake.
\item We construct the Audio Neural Codec Spoof (ANC-Spoof) dataset by applying seven diverse neural codec algorithms to three benchmarks, providing a controlled proxy for codec-induced resynthesis artifacts and enabling systematic evaluation of ADD robustness under this previously unexamined threat model.
\item We propose PCL-NET (Pairwise Consistency Learned Network), a simple training strategy that fine-tunes a pretrained XLS-R (300M) encoder with a pairwise consistency learning objective, discouraging codec-specific artifacts from the real-versus-fake decision boundary.
\end{itemize}
\vspace{-0.2cm}
\section{Literature Review}
In this section, we review prior work on robustness in audio deepfake detection (ADD), moving from conventional signal-level and post-hoc compression manipulations to the more recent line of work on neural codec-based speech generation and resynthesis. 
\vspace{-0.2cm}
\subsection{Robustness in Audio Deepfake Detection}
A substantial number of studies examined how well deepfake detection systems~\cite{uddin2025advbench, farooq2026uncertainty,Khan_2026_CVPR, 11107183, khan2026dual,farooq2025light} hold up against frame or signal-level manipulations applied independently of the generation process. In particular, Wu et al.~\cite{wu2025clad} conducted an early study on the susceptibility of widely adopted ADD systems to such manipulations and proposed CLAD, a contrastive learning-based detection method that minimizes representation variation introduced by these perturbations. Similarly, Tak et al.~\cite{tak2022rawboost} introduced RawBoost, a data augmentation strategy that simulates channel- and codec-like distortions through convolutive and additive noise, without modeling the resynthesis process of any specific codec. Within this broader robustness effort, a smaller set of studies has examined conventional compression as a distortion type. For instance, Li et al.~\cite{li2026measuring} reported that although self-supervised (SSL)-based foundation models perform comparatively well under moderate compression, their performance drops noticeably under extreme conditions, such as low-bitrate MP3 encoding. Klein et al.~\cite{klein2026effect} quantified the impact of telephony transmission on source tracing for audio deepfakes, showing that augmentation strategies such as RawBoost and random audio quantization meaningfully reduce equal error rate (EER) under such conditions. \\
Across this line of work, robustness has been evaluated almost exclusively against post-hoc perturbations, distortions applied to already-generated speech, whether through noise, channel effects, or standard audio codecs such as MP3 or AAC. None of these studies examine distortions introduced by neural audio codecs, whose encoder-decoder architecture and role in modern generation pipelines differ fundamentally from the compression schemes considered above. Whether ADD systems generalize to this distinct distortion type remains an open question, one that we address directly in this work. 

\subsection{Codec-Based Speech Generation and Resynthesis}
A separate and more recent line of work has emerged in response to the growing use of neural codecs as a generative component within audio language model (ALM)-based speech synthesis, rather than as a standalone compression tool. Xie et al.~\cite{xie2025codecfake} introduced Codecfake, a large-scale dataset constructed to support detection of ALM-based synthetic audio. Building on this direction, CodecFake+~\cite{chen2026codecfakeplus} was the first to formally distinguish between Codec Resynthesized (CoRS) speech, real audio passed through a codec's encoder and decoder, and Codec Speech Generation (CoSG), speech generated by an ALM and resynthesized through the same decoder. Because most CoSG systems remain proprietary, CoRS has since been adopted as a practical, accessible proxy for CoSG-style artifacts during training, on the reasoning that both share the same final decoding stage. Building on this proxy relationship, Xiao et al.~\cite{xiao2026how} observed that, precisely because neural codecs serve this dual function, compression and generation, codec-resynthesized data could reasonably be labeled as either bona fide or spoof depending on the intended use case, and studied how this labeling choice affects detection accuracy. More recently, Lin et al.~\cite{lin2025towards} proposed SASTNet, which jointly leverages semantic and acoustic feature encoders to achieve SOTA source-tracing performance specifically on CoSG data.
Taken together, this line of work confirms CoRS as an effective, increasingly standard proxy for CoSG detection and attribution, but it comes with a hidden risk: a detector that learns to associate codec artifacts with the fake class risks misclassifying legitimately compressed real speech. To our knowledge, this risk remains unevaluated. We address this gap by exposing its extent across standard, perturbation-robust, and CoRS-trained ADD systems, and by proposing a codec-aware training strategy that disentangles codec-induced distortion from fakeness itself, improving robustness to legitimate compression while preserving reliable CoSG detection.
\vspace{-0.2cm}
\section{Problem Formulation}
Let $x$ denote a raw speech waveform, and let $y \in \{0,1\}$ denote its ground-truth label. Where $y=0$ corresponds to bona fide speech, and $y=1$ corresponds to deepfake speech. Given a set of $K$ neural audio codecs $\{C_1, C_2, \ldots, C_K\}$, each codec $C_k(\cdot)$ maps a waveform through a neural codec encoder-decoder pipeline to produce a resynthesized waveform $x_c = C_k(x)$. Importantly, codec resynthesis is assumed to be label-preserving. This means that if $x$ is bona fide, $x_c$ is still bona fide speech that has merely undergone lossy compression. Similarly, if $x$ is spoofed, $x_c$ remains spoofed speech. Our goal is to learn a detector $f_\theta(\cdot)$ that produces a classification decision $\hat{y} = f_\theta(x)$ such that $\hat{y}$ remains stable and correct for both $x$ and $x_c$.

\section{Methodology}
\begin{figure*}[t]
    \centering
    \includegraphics[width=\textwidth]{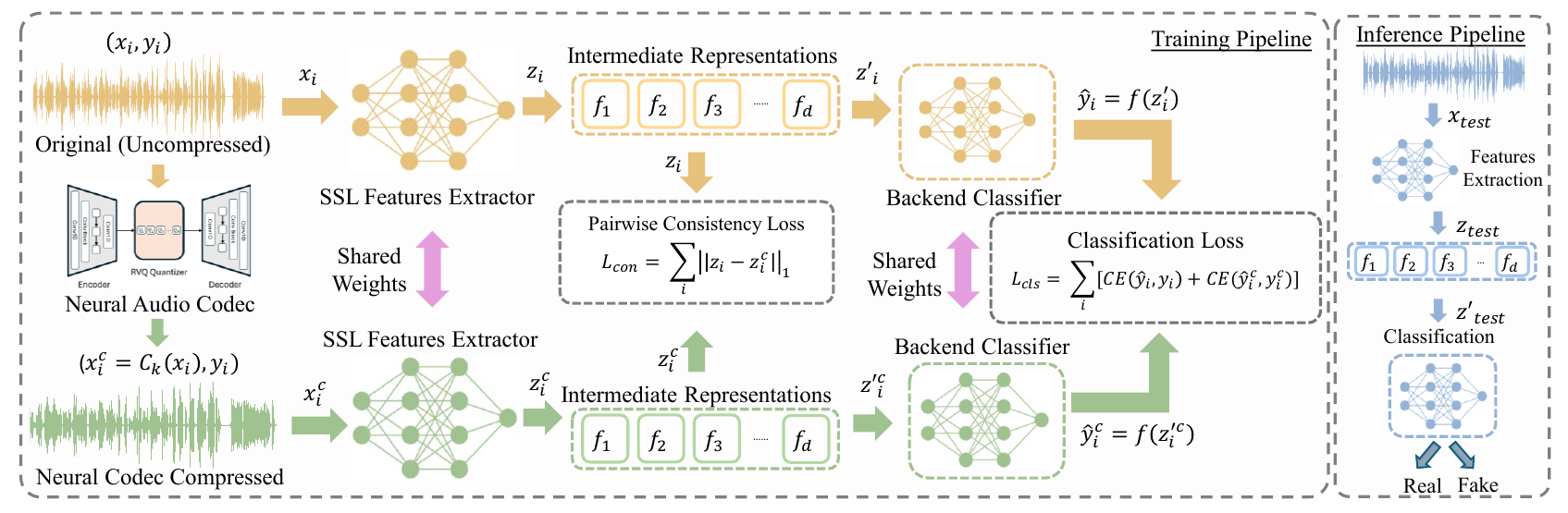}
    \caption{Overview of the proposed training and inference pipeline: SSL features extracted from original and neural-codec-compressed audio are jointly optimized via pairwise consistency and classification losses, with the resulting backend classifier used at inference to distinguish real from fake audio.}
    \vspace{-0.6cm}
    \label{fig:pc_net}
\end{figure*}
\subsection{Overview}
Prior CoRS-based ADD methods use codec resynthesis only as a proxy for codec-based generation, overlooking that the same codecs are also widely used for legitimate compression. In this work, we adopt a simple lightweight training strategy that treats codec resynthesis as a label-preserving augmentation applied to both real and fake speech as shown in Figure~\ref{fig:pc_net}. For this, we introduce a pairwise consistency loss that enforces agreement between an utterance's original and resynthesized SSL embeddings. By explicitly minimizing the representation-level discrepancy between these paired samples, the proposed pairwise consistency learning encourages codec-invariant representations while retaining task-relevant information for deepfake detection. This discourages the model from relying on codec artifacts as a shortcut, making PCL-NET more robust to legitimate compression while preserving competitive performance on codec-based generation.
\subsection{Neural Audio Codecs}
\label{sec:codecs}
To train and evaluate realistic audio neural codec-induced distortions, we employ seven distinct neural audio codecs, with dinstinct configuration discussed as follow:\\[0.3em]
\textbf{BigCodec}~\cite{xin2024bigcodec} scales model capacity to 159M parameters to improve reconstruction at extremely low bitrates around 1 kbps.
 \textbf{ DAC}~\cite{kumar2023highfidelity} is a high-fidelity general-purpose codec that compresses audio to as low as 8 kbps using an improved RVQGAN architecture.
 \textbf{ Higgs Audio V2 Tokenizer}~\cite{higgs_audio_v2_tokenizer} operates at 25 frames per second while matching tokenizers with twice the frame rate. It was trained jointly on speech, music, and sound events.
 \textbf{ Mimi}~\cite{defossez2024moshi} is a streaming codec that compresses 24 kHz audio to 12.5 Hz, distilled from WavLM to preserve semantic content.
 \textbf{ SNAC}~\cite{siuzdak2024snac} extends residual vector quantization with quantizers operating at multiple temporal resolutions for more efficient compression.
 \textbf{ SpeechTokenizer}~\cite{zhang2024speechtokenizer} uses residual vector quantization to disentangle semantic and acoustic information.
 \textbf{ WavTokenizer}~\cite{ji2025wavtokenizer} reconstructs 24 kHz audio using a single quantizer at 40 to 75 tokens per second while maintaining high quality.
 \subsection{Model Architecture}
To tackle the challenge of the audio neural codec compression we present PCL-NET as our proposed baseline method. In particular, PCL-NET, follows a standard SSL-frontend-plus-backend approach commonly used in audio deepfake detection. Specifically, we pair a self-supervised frontend with a lightweight classification backend and add a layer-wise fusion mechanism. For SSL-based feature extraction, we employ the pretrained XLS-R-300M model as our feature extractor, denoted as $\operatorname{Enc}(\cdot)$. Given an input waveform $x$, the frontend produces both a final-layer embedding and the intermediate hidden states from each of its $L$ transformer layers, denoted as $z$. Formally,
\begin{equation}
z = \{f_1,f_2,\ldots,f_L\} = \operatorname{Enc}(x), \quad f_l \in \mathbb{R}^{T' \times d}
\label{eq:enc}
\end{equation}
where $T'$ is the number of frame-level time steps and $d=1024$ is the hidden dimensionality. Each layer's representation $f_l$ is first pooled over time and passed through a shared linear projection with a sigmoid activation to obtain a scalar attention weight $a_l \in (0,1)$. These weights are then used to rescale each layer's full-resolution representation before fusing them across the layer dimension:
\begin{equation}
z' = \sum_{l=1}^{L} a_l \cdot f_l, \quad z' \in \mathbb{R}^{T' \times d}
\label{eq:fusion}
\end{equation}
\begin{table}[t]
\caption{Configurational Details of the Proposed ANC-Spoof Dataset}
\vspace{-0.4cm}
\begin{center}
\small
\begin{tabular}{|c|c|c|c|c|}
\hline
\textbf{Dataset}&\textbf{Split}&\textbf{Orig.}&\textbf{Comp.} & \textbf{Total} \\
\hline
ASV19~\cite{wang2019asvspoof}& Training& 25{,}380& 177{,}660& 203{,}040 \\
\hline
ASV19~\cite{wang2019asvspoof}& Validation& 24{,}986& 174{,}902& 199{,}888 \\
\hline
ASV19~\cite{wang2019asvspoof}& Evaluation& 71{,}933& 503{,}531& 575{,}464 \\
\hline
FoR~\cite{reimao2019for}& Evaluation& 4{,}634& 32{,}438& 37{,}072 \\
\hline
Wild~\cite{muller2022does}& Evaluation& 31{,}780& 222{,}460& 254{,}240 \\
\hline
\multicolumn{2}{|c|}{\textbf{Total}}& \textbf{158{,}713}& \textbf{1{,}110{,}991}& \textbf{1{,}269{,}704} \\
\hline
\multicolumn{5}{l}{\footnotesize $^{\mathrm{*}}$Orig. = uncompressed utterances; Comp. = neural codec-compressed.}
\end{tabular}
\vspace{-0.8cm}
\label{tab:ANC-Spoof}
\end{center}
\end{table}

\begin{table*}[t]
\vspace{-0.8cm}
\caption{Vulnerability of SOTA ADD Systems to Conventional and Neural Codec Compression (EER \%). $^{*}$Standard, $^{\dagger}$Manipulation-robust, $^{\ddagger}$CoRS-based training strategies. Results obtained using author-provided pretrained weights.}
\vspace{-0.7cm}
\begin{center}
\small
\resizebox{\textwidth}{!}{%
\begin{tabular}{|c|c|c|c|c|c|c|c|c|c|c|c|c|c|c|}
\hline
\multirow{2}{*}{\textbf{Model}} & \multirow{2}{*}{\textbf{Dataset}} & \multirow{2}{*}{\textbf{Original}} & \multicolumn{4}{c|}{\textbf{MP3/AAC/Opus}} & \multirow{2}{*}{\textbf{BigCodec}} & \multirow{2}{*}{\textbf{DAC}} & \multirow{2}{*}{\textbf{HiggsV2}} & \multirow{2}{*}{\textbf{Mimi}} & \multirow{2}{*}{\textbf{SNAC}} & \multirow{2}{*}{\textbf{Speech Tokenizer}} & \multirow{2}{*}{\textbf{Wav Tokenizer}} & \multirow{2}{*}{\textbf{Average}} \\
\cline{4-7}
 & & & \textbf{12K} & \textbf{24K} & \textbf{48K} & \textbf{96K} & & & & & & & & \\
\hline
\multirow{3}{*}{AASIST$^{*}$~\cite{jung2022aasist}} & ASV19 & 0.83 & 13.13 & 2.30 & 1.01 & 0.91 & 8.35 & 1.62 & 4.53 & 3.02 & 12.03 & 5.10 & 18.15 & \textbf{7.54} \\
 & Wild & 43.02 & 51.51 & 45.79 & 43.73 & 43.26 & 34.91 & 39.55 & 34.99 & 36.39 & 32.88 & 42.28 & 47.28 & \textbf{38.33} \\
 & FoR & 44.13 & 26.80 & 16.03 & 15.49 & 16.12 & 21.34 & 21.06 & 25.83 & 20.22 & 26.37 & 22.96 & 31.1 & \textbf{24.13} \\
\hline
\multirow{3}{*}{XLS-R-SLS$^{*}$~\cite{zhang2024xlsrsls}} & ASV19 & 0.24 & 7.33 & 3.59 & 0.72 & 0.24 & 4.80 & 0.52 & 1.93 & 1.14 & 13.30 & 3.21 & 18.59 & \textbf{6.21} \\
 & Wild & 7.46 & 27.43 & 13.32 & 9.91 & 9.22 & 25.43 & 10.17 & 17.44 & 13.51 & 37.03 & 27.81 & 46.62 & \textbf{25.43} \\
 & FoR & 10.16 & 38.35 & 23.98 & 14.61 & 10.08 & 25.49 & 19.4 & 49.78 & 24.64 & 58.35 & 40.46 & 51.86 & \textbf{38.57} \\
\hline
\multirow{3}{*}{CLAD$^{\dagger}$~\cite{wu2025clad}} & ASV19 & 1.11 & 4.09 & 2.27 & 1.23 & 1.20 & 7.26 & 1.69 & 4.05 & 2.81 & 11.68 & 4.52 & 16.56 & \textbf{6.94} \\
 & Wild & 36.35 & 43.62 & 40.53 & 39.11 & 38.57 & 33.62 & 34.1 & 33.93 & 35.17 & 39.91 & 37.75 & 34.71 & \textbf{35.6} \\
 & FoR & 40.76 & 49.78 & 38.48 & 42.02 & 43.76 & 44.48 & 48.32 & 49.12 & 44.43 & 55.16 & 45.77 & 43.42 & \textbf{47.24} \\
\hline
\multirow{3}{*}{RawNet2-RawBoost$^{\dagger}$~\cite{tak2022rawboost}} & ASV19 & 3.81 & 4.42 & 3.79 & 3.75 & 3.77 & 10.29 & 4.72 & 7.71 & 5.86 & 11.44 & 8.2 & 16.71 & \textbf{9.28} \\
 & Wild & 37.48 & 41.03 & 39.94 & 40.05 & 40.09 & 35.99 & 35.26 & 36.08 & 37.41 & 35.65 & 35.29 & 35.92 & \textbf{35.94} \\
 & FoR & 38.95 & 42.66 & 38.15 & 37.14 & 36.71 & 43.46 & 43.38 & 45.53 & 43.59 & 47.84 & 44.04 & 51.86 & \textbf{45.67} \\
\hline
\multirow{3}{*}{W2W2-AASIST$^{\ddagger}$~\cite{xie2025codecfake}} & ASV19 & 3.81 & 14.87 & 6.21 & 6.38 & 3.90 & 18.88 & 7.13 & 14.25 & 8.76 & 25.42 & 17.89 & 30.77 & \textbf{17.59} \\
 & Wild & 9.61 & 18.36 & 12.67 & 10.23 & 9.97 & 21.65 & 14.53 & 22.79 & 15.91 & 32.13 & 30.98 & 28.45 & \textbf{23.78} \\
 & FoR & 0.67 & 25.65 & 14.45 & 3.65 & 1.45 & 19.53 & 13.21 & 51.14 & 19.4 & 62.06 & 34.1 & 40.72 & \textbf{34.31} \\
\hline
\multirow{3}{*}{MMS-300M$^{\ddagger}$~\cite{ge2025posttraining}} & ASV19 & 0.69 & 18.34 & 4.88 & 1.74 & 0.87 & 14.52 & 15.72 & 33.09 & 34.23 & 38.36 & 36.27 & 34.10 & \textbf{29.47} \\
 & Wild & 2.50 & 19.25 & 3.99 & 2.61 & 2.45 & 33.70 & 14.24 & 29.16 & 30.37 & 58.47 & 50.81 & 52.28 & \textbf{38.43} \\
 & FoR & 2.31 & 23.63 & 8.65 & 3.56 & 1.90 & 48.57 & 19.12 & 28.74 & 23.15 & 19.44 & 32.78 & 27.34 & \textbf{28.45} \\
\hline
\multirow{3}{*}{XLS-R-2B$^{\ddagger}$~\cite{ge2025posttraining}} & ASV19 & 0.15 & 15.72 & 4.28 & 0.59 & 0.22 & 20.38 & 10.25 & 29.96 & 30.02 & 43.63 & 38.6 & 38.76 & \textbf{30.23} \\
 & Wild & 1.32 & 6.18 & 1.71 & 1.5 & 1.37 & 32.07 & 7.57 & 23.58 & 12.65 & 43.57 & 33.98 & 40.96 & \textbf{27.77} \\
 & FoR & 1.01 & 35.67 & 9.8 & 3.71 & 1.06 & 59.37 & 29.22 & 59.91 & 34.14 & 56.02 & 54.94 & 64.83 & \textbf{51.2} \\
\hline
\end{tabular}}
\label{tab:sota_vuln}
\end{center}
\end{table*}
We use $z'$, the fused representation, as input to the backend. The backend classifier consists of batch normalization followed by a SELU activation and a temporal max-pooling layer. This classifier projects the features into a hidden space and outputs two class logits via a log-softmax function. The core idea behind PCL-NET is to prevent the model from latching onto codec-induced distortions as a shortcut for deepfake detection. For a training utterance $x_i$ with label $y_i$, we generate a compressed version $x_i^{c} = C_k(x_i)$ using a neural codec $C_k$. We then extract embeddings $z_i$ and $z_i^{c}$ from the shared encoder pipeline and minimize their $L_1$ distance over $N$ utterances:
\begin{equation}
\mathcal{L}_{\mathrm{con}}
=
\frac{1}{N}
\sum_{i=1}^{N}
\left\| z_i - z_i^{c} \right\|_{1}
\label{eq:con}
\end{equation}
We favor $L_1$ distance over contrastive objectives as it directly enforces strict point-to-point feature invariance across compressed views without requiring negative sample mining. This pushes codec-induced variation out of the representation space relied on by the classifier. Notably, the consistency constraint operates purely at the loss level. PCL-NET uses a single encoder and classifier, where the pair $(x_i, x_i^{c})$ is only formed during training to compute this regularization signal. We train the model jointly using a standard cross-entropy classification loss $\mathcal{L}_{\mathrm{cls}}$ applied to both the original and codec-compressed views, combined with the consistency loss:
\begin{equation}
\mathcal{L} = \mathcal{L}_{\mathrm{cls}} + \lambda  \mathcal{L}_{\mathrm{con}}
\label{eq:total_loss}
\end{equation}
where $\lambda = 0.7$ balances the two objectives. Because pairing is only required to compute $\mathcal{L}_{\mathrm{con}}$ during training, inference requires only a single forward pass of an unseen utterance.
\vspace{-0.2cm}
\begin{table*}[!t]
\vspace{-0.7cm}
\caption{Robustness of PCL-NET Trained with Different Neural Codecs for Consistency Learning (EER \%).}
\vspace{-0.7cm}
\begin{center}
\small
\resizebox{\textwidth}{!}{%
\begin{tabular}{|c|c|c|c|c|c|c|c|c|c|c|c|c|c|c|}
\hline
\multirow{2}{*}{\textbf{Codec Method(s)}} & \multirow{2}{*}{\textbf{Dataset}} & \multirow{2}{*}{\textbf{Original}} & \multicolumn{4}{c|}{\textbf{MP3/AAC/Opus}} & \multirow{2}{*}{\textbf{BigCodec}} & \multirow{2}{*}{\textbf{DAC}} & \multirow{2}{*}{\textbf{HiggsV2}} & \multirow{2}{*}{\textbf{Mimi}} & \multirow{2}{*}{\textbf{SNAC}} & \multirow{2}{*}{\textbf{Speech Tokenizer}} & \multirow{2}{*}{\textbf{Wav Tokenizer}} & \multirow{2}{*}{\textbf{Average}} \\
\cline{4-7}
 & & & \textbf{12K} & \textbf{24K} & \textbf{48K} & \textbf{96K} & & & & & & & & \\
\hline
\multirow{3}{*}{Baseline} & ASV19 & 0.34 & 7.97 & 5.65 & 0.83 & 0.38 & 5.11 & 0.75 & 2.35 & 1.74 & 13.39 & 4.02 & 21.10 & \textbf{6.92} \\
 & Wild & 8.28 & 25.13 & 10.94 & 8.02 & 7.75 & 23.68 & 8.65 & 15.31 & 12.05 & 36.87 & 24.84 & 43.31 & \textbf{23.53} \\
 & FoR & 11.35 & 35.78 & 24.47 & 14.48 & 9.58 & 31.77 & 22.49 & 49.50 & 32.41 & 61.31 & 42.58 & 53.45 & \textbf{41.93} \\
\hline
\multirow{3}{*}{BigCodec~\cite{xin2024bigcodec}} & ASV19 & 0.14 & 5.98 & 1.28 & 0.23 & 0.19 & 2.55 & 0.41 & 1.44 & 1.11 & 12.67 & 2.88 & 15.25 & \textbf{5.19} \\
 & Wild & 9.29 & 27.01 & 12.67 & 10.45 & 10.10 & 17.56 & 9.49 & 14.54 & 12.91 & 26.63 & 20.86 & 32.46 & \textbf{19.21} \\
 & FoR & 5.52 & 17.67 & 10.42 & 6.41 & 5.39 & 22.62 & 12.49 & 39.62 & 15.24 & 45.62 & 30.71 & 46.33 & \textbf{30.38} \\
\hline
\multirow{3}{*}{DAC~\cite{kumar2023highfidelity}} & ASV19 & 0.16 & 9.23 & 3.98 & 0.46 & 0.16 & 4.57 & 0.34 & 1.61 & 0.98 & 15.01 & 3.18 & 19.59 & \textbf{6.47} \\
 & Wild & 8.50 & 25.30 & 11.77 & 8.94 & 8.40 & 23.32 & 9.12 & 15.87 & 11.98 & 32.30 & 24.72 & 42.38 & \textbf{22.81} \\
 & FoR & 7.86 & 39.19 & 24.13 & 12.32 & 6.54 & 28.89 & 17.26 & 49.33 & 22.87 & 58.74 & 39.45 & 53.04 & \textbf{38.51} \\
\hline
\multirow{3}{*}{HiggsV2~\cite{higgs_audio_v2_tokenizer}} & ASV19 & 0.21 & 7.19 & 2.24 & 0.22 & 0.18 & 4.08 & 0.30 & 0.74 & 0.31 & 12.98 & 1.85 & 17.47 & \textbf{5.39} \\
 & Wild & 7.28 & 22.45 & 9.53 & 7.79 & 7.49 & 20.19 & 8.06 & 13.24 & 10.11 & 27.00 & 21.12 & 35.68 & \textbf{19.34} \\
 & FoR & 5.65 & 31.14 & 17.54 & 8.35 & 5.16 & 21.11 & 12.58 & 35.99 & 16.16 & 55.18 & 27.13 & 48.36 & \textbf{30.93} \\
\hline
\multirow{3}{*}{Mimi~\cite{defossez2024moshi}} & ASV19 & 0.19 & 8.16 & 2.88 & 0.27 & 0.19 & 4.69 & 0.26 & 1.03 & 0.54 & 14.00 & 2.03 & 18.70 & \textbf{5.89} \\
 & Wild & 7.54 & 21.59 & 9.86 & 8.00 & 7.71 & 22.70 & 8.12 & 14.60 & 10.40 & 28.70 & 22.50 & 39.30 & \textbf{20.90} \\
 & FoR & 8.83 & 36.17 & 21.99 & 11.85 & 7.64 & 29.54 & 17.72 & 45.77 & 22.40 & 57.38 & 35.82 & 51.86 & \textbf{37.21} \\
\hline
\multirow{3}{*}{SNAC~\cite{siuzdak2024snac}} & ASV19 & 0.72 & 3.71 & 0.91 & 0.72 & 0.70 & 5.84 & 0.97 & 1.78 & 1.45 & 4.84 & 2.65 & 15.62 & \textbf{4.74} \\
 & Wild & 14.74 & 26.24 & 16.12 & 15.26 & 15.38 & 24.62 & 16.07 & 21.09 & 18.06 & 24.98 & 23.81 & 32.78 & \textbf{23.06} \\
 & FoR & 7.94 & 16.88 & 10.77 & 7.25 & 6.54 & 20.09 & 12.37 & 22.16 & 17.18 & 26.72 & 21.23 & 37.29 & \textbf{22.43} \\
\hline
\multirow{3}{*}{Speech Tokenizer~\cite{zhang2024speechtokenizer}} & ASV19 & 0.27 & 6.71 & 1.63 & 0.22 & 0.25 & 4.41 & 0.25 & 0.75 & 0.53 & 12.40 & 1.28 & 16.80 & \textbf{5.20} \\
 & Wild & 7.76 & 24.42 & 10.16 & 8.29 & 7.92 & 19.90 & 8.01 & 13.30 & 10.40 & 26.50 & 20.80 & 36.90 & \textbf{19.40} \\
 & FoR & 7.34 & 29.78 & 17.67 & 8.22 & 6.13 & 22.31 & 13.92 & 33.43 & 17.98 & 52.52 & 24.08 & 45.90 & \textbf{30.02} \\ \hline
\multirow{3}{*}{Wav Tokenizer~\cite{ji2025wavtokenizer}} & ASV19 & 1.58 & 5.40 & 2.58 & 1.87 & 1.82 & 6.52 & 2.19 & 3.51 & 2.46 & 10.21 & 5.78 & 12.40 & \textbf{6.15} \\
 & Wild & 21.92 & 39.37 & 26.99 & 24.55 & 24.14 & 24.29 & 17.06 & 22.67 & 20.21 & 30.29 & 24.57 & 38.79 & \textbf{25.41} \\
 & FoR & 6.67 & 13.88 & 8.70 & 6.71 & 6.09 & 19.08 & 11.98 & 26.41 & 14.72 & 38.48 & 22.49 & 36.75 & \textbf{24.27} \\ \hline
\multirow{3}{*}{All} & ASV19 & 1.13 & 3.14 & 2.03 & 1.39 & 1.22 & 5.67 & 1.75 & 3.08 & 2.65 & 5.92 & 4.23 & 12.77 & \textbf{5.15} \\
 & Wild & 18.04 & 25.56 & 18.75 & 17.97 & 18.00 & 23.65 & 17.28 & 21.92 & 18.50 & 25.28 & 23.75 & 33.67 & \textbf{23.44} \\
 & FoR & 2.24 & 7.38 & 4.51 & 2.74 & 2.24 & 7.64 & 3.32 & 9.58 & 4.23 & 10.96 & 7.81 & 25.23 & \textbf{9.82} \\
\hline
\end{tabular}}
\vspace{-0.8cm}
\label{tab:pcnet}
\end{center}
\end{table*}
\section{Experimental Results}
This section presents the vulnerability of SOTA ADD systems to neural codec compression and the development of ANC-Spoof, our neural codec-compressed dataset, and the performance analysis of SOTA and proposed PCL-NET.
\subsection{Dataset Details}
We construct the ANC-Spoof dataset by applying neural codec compression on top of the existing three ADD benchmark datasets. These datasets include ASVspoof 2019 LA~\cite{wang2019asvspoof}, In-the-Wild~\cite{muller2022does}, and FoR~\cite{reimao2019for}. The training and validation splits of ASVspoof 2019 LA \cite{wang2019asvspoof} are reserved for model development, while the evaluation partitions of all three datasets are reserved for generalization testing. For every utterance in these splits, we generate seven codec-compressed counterparts using the neural codecs described in Section~\ref{sec:codecs}, applying compression to both real and fake samples. The ground-truth labels of both real and fake are preserved under compression.
Table~\ref{tab:ANC-Spoof} summarizes the resulting dataset. Altogether, ANC-Spoof contains 158,713 original utterances and 1,110,991 codec-compressed utterances (seven codecs applied to each original utterance per split), bringing the total to 1,269,704 utterances across training, validation, and evaluation. Across all our experiments, we report performance in terms of Equal Error Rate (EER). Beyond ANC-Spoof, we also evaluate on two additional benchmarks, CodecFake+~\cite{chen2026codecfakeplus} and CodecDeepfakeDetection (CDD)~\cite{xiao2026how}. Here, we restrict evaluation to their real and CoSG samples to verify that our codec-aware training does not come at the cost of detection performance on CoSG-based deepfakes.

\subsection{ADD Vulnerability to Neural Codec Compression}
\label{sec:sota_vuln}
In this experiment, we evaluate and expose seven SOTA ADD systems on our proposed ANC-Spoof dataset, covering standard, manipulation-robust, and CoRS-based training strategies, to assess the robustness and performance analysis as how existing detectors hold up under compression. For each utterance, we test performance under all seven neural codecs, as well as conventional compression (MP3/AAC/Opus at 12K, 24K, 48K, and 96K bitrates, with one method randomly sampled per utterance per bitrate). Table~\ref{tab:sota_vuln} reports the resulting EERs.
While low-bitrate conventional codecs (e.g., 12K) degrade performance, neural codecs cause far more severe performance drops across all systems. Non-CoRS systems (AASIST \cite{jung2022aasist}, XLS-R-SLS \cite{zhang2024xlsrsls}, CLAD \cite{wu2025clad}, RawNet2-RawBoost \cite{tak2022rawboost}) show a substantial drop under codec compression, though the degradation stays comparatively moderate. CoRS-trained systems (W2W2-AASIST~\cite{xie2025codecfake}, MMS-300M~\cite{ge2025posttraining}, XLS-R-2B~\cite{ge2025posttraining}), which use codec-resynthesized data as a proxy for CoSG during training, degrade far more severely. XLS-R-2B \cite{ge2025posttraining}, for instance, achieves just 1.01\% EER on original FoR \cite{reimao2019for} audio, yet its average EER rises to 51.2\% under codec compression, near chance level. This confirms our hypothesis that CoRS-based training causes models to associate codec artifacts with the fake class, which is consistent with a higher likelihood of misclassifying genuinely compressed bona fide speech as fake. \vspace{-0.3cm}
\subsection{Robustness of PCL-NET via Consistency Learning}
In this experiment, we evaluate PCL-NET and analyze how the choice of training codec affects robustness. For each codec, we train a separate PCL-NET variant using only that codec for consistency learning. We also include a \textit{Baseline} (no consistency loss, trained on uncompressed audio) and an \textit{All} variant (randomly selecting one of the seven codecs per utterance). Table~\ref{tab:pcnet} reports the results. 
Compared to \textit{Baseline}, consistency-trained variants substantially reduce EER under neural compression while also improving robustness across conventional codecs (e.g., \textit{All} drops FoR 12K EER from 35.78\% to 7.38\%). Codecs designed for speech compression (BigCodec \cite{xin2024bigcodec}, HiggsV2 \cite{higgs_audio_v2_tokenizer}, SNAC \cite{siuzdak2024snac}) yield stronger gains than CoSG-focused tokenizers (SpeechTokenizer \cite{zhang2024speechtokenizer}, WavTokenizer \cite{ji2025wavtokenizer}). Crucially, \textit{All} achieves the best overall balance, reducing average neural codec EER to 5.15\%, 23.44\%, and 9.82\% on ASV19 \cite{wang2019asvspoof}, Wild \cite{muller2022does}, and FoR \cite{reimao2019for}. This confirms that consistency learning generalizes  when exposed to diverse neural codecs.  \vspace{-0.3cm}
 \subsection{Generalization to CoSG Detection}
 After robustness analysis of the SOTA and proposed PCL-NET, in this experiment we evaluated the generalization of the ADD systems on CodecFake+~\cite{chen2026codecfakeplus} and CDD~\cite{xiao2026how}, restricted to real and CoSG samples, to check whether codec-aware training affects CoSG detection. Since CoRS-trained systems (W2W2-AASIST \cite{xie2025codecfake}, MMS-300M \cite{ge2025posttraining}, XLS-R-2B \cite{ge2025posttraining}) are explicitly optimized for this setting, our goal is not to surpass them, but to verify that PCL-NET's improved compression robustness does not introduce a counter-effect on CoSG detection. As Table~\ref{tab:cosg_generalization} shows, CoRS-trained systems perform strongly here, consistent with their training objective. However, this comes at the cost of high vulnerability to legitimate compression shown in Section~\ref{sec:sota_vuln}. In contrast, our codec-aware variants substantially outperform non-CoRS baselines on CoSG detection while remaining robust to legitimate compression, striking a more favorable balance. \vspace{-0.4cm}
\section{Conclusion}
 \vspace{-0.15cm}
In this work, we exposed the vulnerability of existing ADD systems to legitimate neural codec compression, an increasingly relevant threat. We introduced ANC-Spoof, built with seven neural codecs across three ADD benchmarks, and showed that all evaluated detectors, especially CoRS-trained models, suffer substantial performance degradation, often misclassifying compressed bona fide speech as fake. To improve robustness, we proposed PCL-NET, which combines codec-aware augmentation with a pairwise consistency objective to learn codec-invariant representations while preserving CoSG detection performance. A limitation is that ANC-Spoof uses post-hoc codec compression rather than native ALM-generated CoSG samples, which may contain additional token-prediction artifacts. Future work should evaluate robustness on native CoSG data as such systems become more accessible.
\vspace{-0.23cm}
\begin{table}[t]
\caption{Generalization to CoSG Detection (EER \%).}
\vspace{-0.4cm}
\begin{center}
\small
\begin{tabular*}{\columnwidth}{@{\extracolsep{\fill}}|l|c|c|}
\hline
\textbf{Model} & \textbf{CodecFake+~\cite{chen2026codecfakeplus}} & \textbf{CDD~\cite{xiao2026how}} \\
\hline
AASIST~\cite{jung2022aasist} & 42.23 & 37.72 \\
\hline
XLS-R-SLS~\cite{zhang2024xlsrsls} & 24.26 & 31.32 \\
\hline
CLAD~\cite{wu2025clad} & 44.57 & 41.95 \\
\hline
RawNet2~\cite{tak2022rawboost} & 44.9 & 40.60 \\
\hline
W2W2-AASIST~\cite{xie2025codecfake} & 9.9 & 16.89 \\
\hline
MMS-300M~\cite{ge2025posttraining} & 5.5 & 6.51 \\
\hline
XLS-R-2B~\cite{ge2025posttraining} & 3.67 & 3.64 \\
\hline
Baseline (Ours) & 21.81 & 29.39 \\
\hline
BigCodec (Ours) & 22.04 & 24.29 \\
\hline
HiggsV2 (Ours) & 22.65 & 26.35 \\
\hline
SNAC (Ours) & 31.28 & 24.65 \\
\hline
All (Ours) & 31.28 & 23.09 \\
\hline
\end{tabular*}
\label{tab:cosg_generalization}
\vspace{-0.8cm}
\end{center}
\end{table}
\section*{Acknowledgment}
This work was supported by NSF Grant No. 2409577; the views expressed are those of the authors and do not necessarily reflect those of the NSF.
\vspace{-0.1cm}

\bibliographystyle{IEEEtran}
\bibliography{mybib}
\end{document}